\documentclass[]{spie} 
 
\usepackage{amsmath,amsfonts,amssymb}
\usepackage{graphicx}
\usepackage[colorlinks=true, allcolors=blue]{hyperref}
\usepackage{multirow}

\title{Broadband Vector Vortex Coronagraph Testing at NASA’s High Contrast Imaging Testbed Facility}

\author[a]{Garreth Ruane}
\author[a]{A J Eldorado Riggs}
\author[a]{Eugene Serabyn}
\author[a]{Wesley Baxter}
\author[a]{Camilo Mejia~Prada}
\author[b]{Dimitri Mawet}
\author[a]{Matthew Noyes}
\author[a]{Phillip K. Poon}
\author[c]{Nelson Tabiryan}
\affil[a]{Jet Propulsion Laboratory, California Institute of Technology, 4800 Oak Grove Dr., Pasadena, CA 91109}
\affil[b]{Department of Astronomy, California Institute of Technology, 1200 E. California Blvd., Pasadena, CA 91125}
\affil[c]{Beam Engineering for Advanced Measurements Co., 1300 Lee Rd., Orlando, FL 32810}

\authorinfo{Send correspondence to gruane@jpl.nasa.gov.}

\begin{document} 
\maketitle

\begin{abstract}
The unparalleled theoretical performance of an ideal vector vortex coronagraph makes it one of the most promising technologies for directly imaging exoplanets with a future, off-axis space telescope. However, the image contrast required for observing the light reflected from Earth-sized planets ($\sim10^{-10}$) has yet to be demonstrated in a laboratory setting. With recent advances in the manufacturing of liquid crystal vector vortex waveplates as well as system-level performance improvements on our testbeds, we have achieved raw contrast of 1.6$\times10^{-9}$ and 5.9$\times10^{-9}$ in 10\% and 20\% optical bandwidths, respectively, averaged over 3-10~$\lambda/D$ separations on one side of the pseudo-star. The former represents a factor of 10 improvement over the previously reported performance. We show experimental comparisons of the contrast achieved as a function of spectral bandwidth. We provide estimates of the limiting error terms and discuss the improvements needed to close the gap in contrast performance required for future exoplanet imaging space telescopes.
\end{abstract}


\section{Introduction}
\label{sec:intro} 

Exoplanet imaging and spectroscopy with a coronagraph instrument is a key capability for NASA's next generation space telescope\cite{Astro2020,HabEx_finalReport,LUVOIR_finalReport}. Since a typical Solar-type star is $\sim$10 orders of magnitude brighter than the reflected light from a temperate, Earth-sized planet orbiting it, the coronagraph instrument is designed to precisely suppress the light from the star that would otherwise dominate the faint exoplanet at a projected angular separation of only a fraction of an arcsecond away. The primary goal of the coronagraph is to reduce the noise due to the star below that of the exoplanet signal by minimizing the starlight in a region next to the star, known as the ``dark hole" or ``dark zone," over a significant spectral bandwidth ($\Delta\lambda/\lambda\approx$10\%-20\%) while maintaining throughput for exoplanets that are separated from the star by only a few times the angular resolution of the telescope ($\lambda/D$). 

The main technical challenge is to readily achieve sufficient ``raw contrast" on the order of $10^{-10}$ within the largest possible spectral bandwidth and field of view (typically extending on the order of 1 arcsecond from the star). The raw contrast is formally defined by the residual intensity due to the on-axis point source (i.e. the star) at an off-axis location normalized by the signal that would be seen from a point source of equal brightness at the same location in the image. The best raw contrast performance achieved to date is $4\times10^{-10}$ for a simple Lyot coronagraph in a 10\% bandwidth with central wavelength of $\lambda_0$~=~550~nm averaged over a 3-10 $\lambda_0/D$ annular region around the star\cite{Seo2019}. While the demonstrated contrast may be sufficiently close to the $10^{-10}$ goal, the Lyot coronagraph design degrades planet throughput significantly due to the strong apodization required to create the dark zone and is relatively sensitive to low-order aberrations leading to tight wavefront stability requirements. 

Vortex coronagraphs\cite{Mawet2005,Foo2005} are a promising alternative to Lyot coronagraphs for unobscured telescopes due to their theoretical high throughput, broad bandwidth, large field of view, and insensitivity to low-order aberrations\cite{Ruane2018_JATIS}. To date, the best broadband contrast performance has been achieved in the laboratory with a multi-layer liquid crystal implementation of the focal plane mask\cite{Mawet2009,Serabyn2019} which has been tested on various coronagraph testbeds\cite{Serabyn2013,SerabynTDEM1,SerabynTDEM2,MejiaPrada2019,LlopSayson2020,Ruane2020_SPIE,Riggs2021_SPIE}. In the previous installment of these proceedings, our team reported $10^{-8}$ contrast in a 10\% bandwidth\cite{Ruane2020_SPIE}. Here, we show an order of magnitude improvement on these previously reported results and discuss the pathway to closing the contrast gap between vortex and Lyot coronagraphs on our testbeds. 

\section{The focal plane mask}

In this section, we present the theoretical design and laboratory characterization of the vector vortex waveplate used as the phase-only focal plane mask in our vortex coronagraph laboratory demonstration described below.  

\subsection{Mask design}

A vector vortex coronagraph mask is a diffractive waveplate equivalent to an achromatic half-wave plate (HWP) with a spatially-varying fast axis. For a fast axis angle $\chi$, the Jones matrix in the circular polarization basis is\cite{Ruane2020_SPIE}
\begin{equation}
    \mathbf{M}_\circlearrowright = 
    \left[ \begin{matrix}
        0 & e^{i2\chi}  \\
        e^{-i2\chi} & 0  \\
    \end{matrix} \right].
    \label{eqn:jones1}
\end{equation}
Patterning the fast axis angle, $\chi$, according to an arbitrary two-dimensional function imparts a phase shift of $\pm$2$\chi$ on the transmitted beam, the sign of which depends on the handedness of the incident circular polarization. The fast axis pattern $\chi=l\theta/2$, where $l$ is an integer known as the charge and $\theta$ is the azimuthal angle, produces a vortex phase pattern at the output: $\exp(\pm i l \theta)$.

To achieve the vortex phase at all wavelengths within the desired spectral band, BEAM Engineering\footnote{\url{https://www.beamco.com/}} has employed a technique that uses three uniform layers of birefringent material, where each layer is made up of a few-hundred-nanometer thick liquid crystal polymer (LCP) with fast axis angles that are offset by 60$^{\circ}$\cite{Pancharatnam1955,Koester1959}. The result is an achromatic HWP in the form of a $\sim$1~$\mu$m thick LCP slab that is contained between two glass substrates with anti-reflective coatings\cite{Serabyn2019}.

The mask manufactured for this study was received by JPL in Sept. 2020. The design is an achromatic charge $l$~=~4 vortex mask for a bandwidth of at least $\Delta\lambda/\lambda$~=~0.1 centered at 650~nm. The central phase singularity is blocked by a circular occultor 15~$\mu$m in diameter. 

\subsection{Mask characterization} 

We characterized the vortex mask using a Mueller matrix spectro-polarimeter (MMSP; Axometrics Axostep) that measures the Mueller matrix (MM) for transmitted light at each pixel in a microscope image of the mask at several visible wavelengths with diffuse transmitted illumination. In our setup, the MM was measured at wavelengths 450-800~nm in sequential steps of 50~nm at a spatial sampling of 5.8~$\mu$m per pixel for 128$\times$160 pixels (or 0.74$\times$0.93~mm).

\begin{figure}[t]
    \centering
    \includegraphics[width=0.8\linewidth]{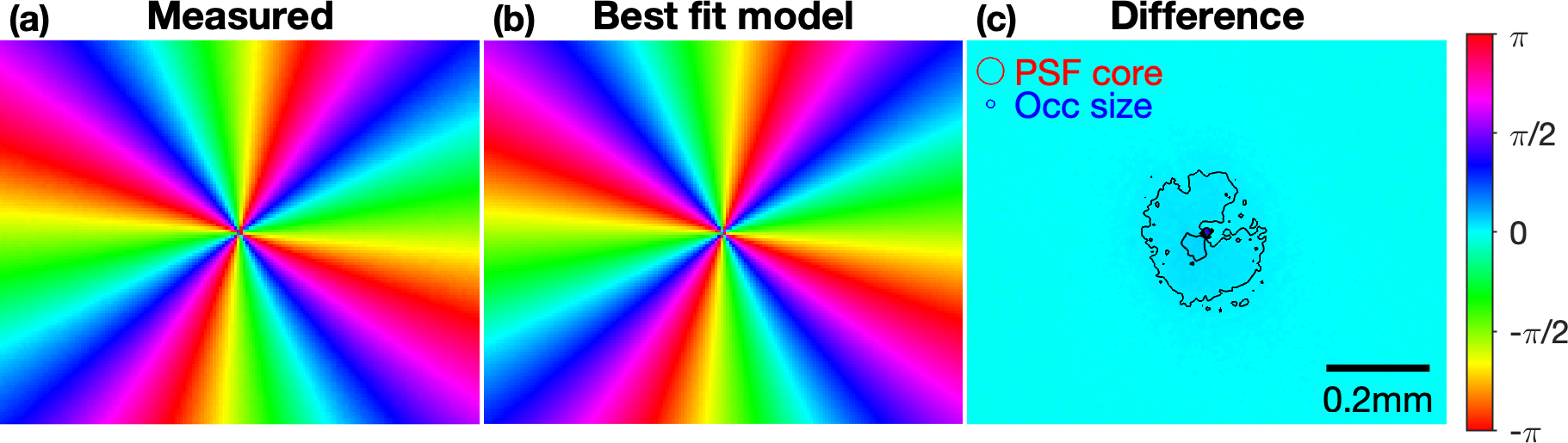}
    \caption{Focal plane mask phase pattern characterization at $\lambda$~=~650~nm. \textbf{(a)}~Geometric phase shift based on the measured fast axis orientation. \textbf{(b)}~The best fit model of the vortex pattern. \textbf{(c)}~The difference between (a) and (b). The red circle shows the size of the point spread function (PSF) core in the testbed configuration presented below (i.e. an angular radius of 1.22$\lambda/D$ corresponding to the first null in the Airy pattern). The central occultor (blue circle) is 15~$\mu$m in diameter. The contour lines show phase steps of $\pi$/40. }
    \label{fig:AxoStep_phz}
\end{figure}

Figure~\ref{fig:AxoStep_phz}a shows the phase shift for one circular polarization based on the measured orientation of the fast axis, $\chi$, at $\lambda$~=~650~nm. We fit the expected $\exp(i 4 \theta)$ phase pattern to the measurement to numerically determine the vortex center and azimuthal rotation angle offset (see Fig.~\ref{fig:AxoStep_phz}b). When measuring the fast axis orientation, the MMSP has an ambiguity where any $\chi$=$\chi+n\pi$ radians is also valid. Thus, we unwrapped the fast axis angles by adopting the $\pi$-offset that gives a phase shift that best matches the vortex pattern at each pixel. Then, we subtracted the best fit vortex pattern to determine the errors in the phase pattern due to fast axis orientation erorrs (see Fig.~\ref{fig:AxoStep_phz}c). We find that the phase error is centrally peaked and slightly asymmetric with a peak of $\sim\pi$/10. The first $\pi$/40 contour occurs at approximately 122~$\mu$m from the center and $\pi$/20 is exceeded within the inner southwest lobe at a maximum radius of 52.3~$\mu$m. 

\begin{figure}[t]
    \centering
    \includegraphics[height=0.36\linewidth]{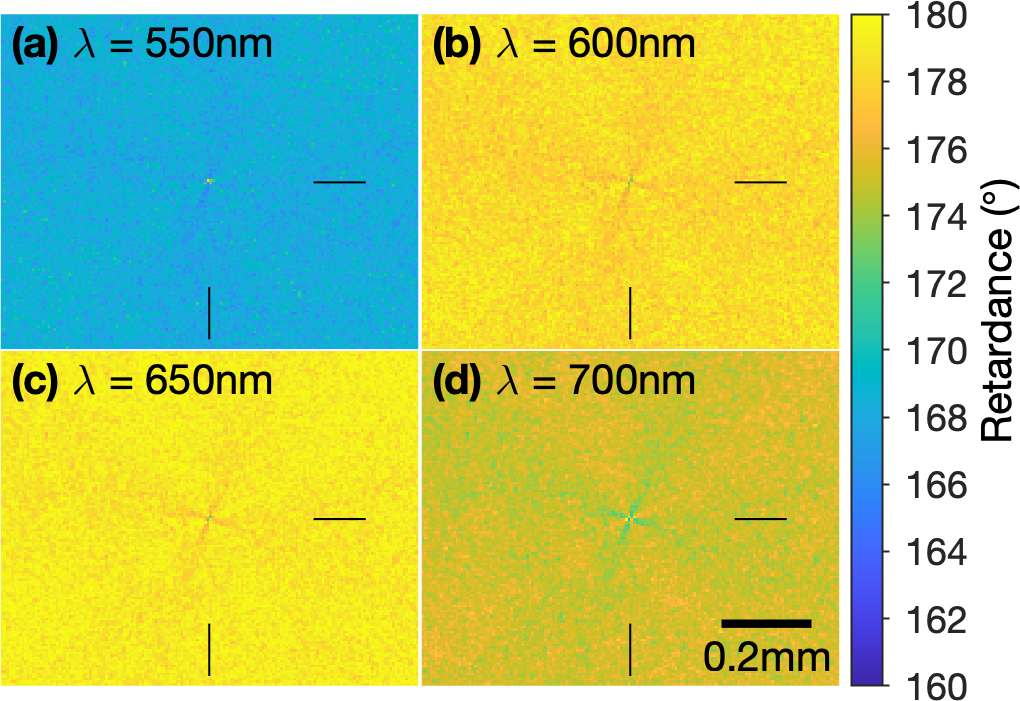}\hfill
    \includegraphics[height=0.36\linewidth]{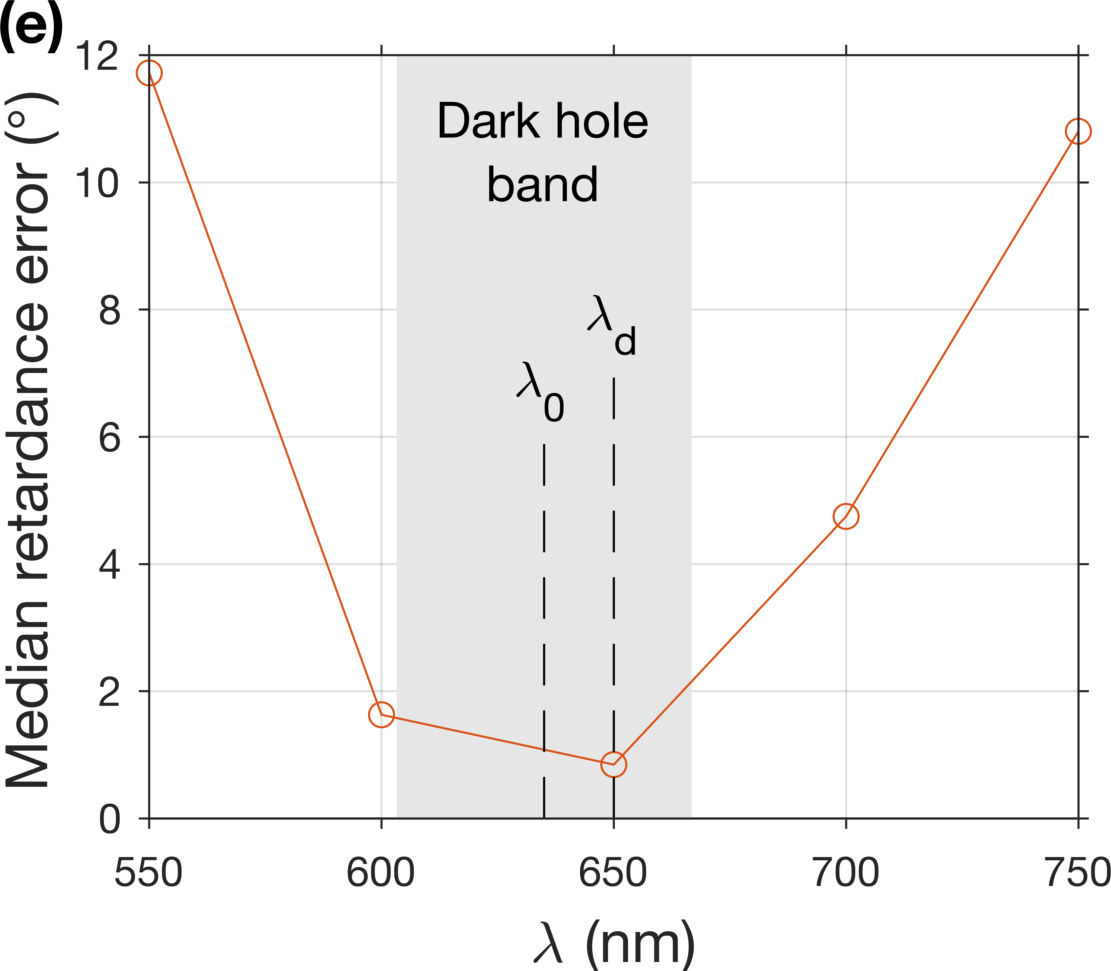}
    \caption{Focal plane mask retardance measurements. \textbf{(a)-(d)}~Retardance maps at wavelengths of (a)~550~nm, (b)~600~nm, (c)~650~nm, and (d)~700~nm. \textbf{(e)}~Median retardance error within images shown in (a)-(d). The gray area indicates the wavelength range used for the testbed results below, which has a central wavelength $\lambda_0$~=~635~nm and bandwidth of $\Delta\lambda/\lambda$~=~0.1. The design wavelength of the mask is $\lambda_d$~=~650~nm. }
    \label{fig:AxoStep_ret}
\end{figure}

The second quantity of interest that we measured with the MMSP is the retardance. Figure~\ref{fig:AxoStep_ret}a-d show the retardance maps at four visible wavelengths: 550, 600, 650, and 700~nm. While the vortex mask would ideally have 180$^\circ$ retardance for all positions and wavelengths, the manufactured masks show a global bias in the retardance that depends strongly on the wavelength. In addition, there is a cross-like defect near the center of the mask at all wavelengths (indicated by fiducial hash marks) as well as small variations at high spatial frequencies. The median retardance error (Fig~\ref{fig:AxoStep_ret}e) is $<$2$^\circ$ within a $\Delta\lambda/\lambda$~=~0.1 passband centered around $\lambda_0$~=~635~nm. In the following, we use $\lambda_0$~=~635~nm as the central wavelength, which is slightly smaller that the design wavelength of $\lambda_d$~=~650~nm, to target the region with the lowest bulk retardance error. Table~\ref{tab:ret_err} provides the retardance error statistics determined from the MMSP images. 

\begin{table}[t]
\caption{Retardance error statistics in images shown in Fig.~\ref{fig:AxoStep_ret}a-d.} 
\label{tab:ret_err}
\begin{center}       
\begin{tabular}{|c|c|c|c|c|} 
\hline
\rule[-1ex]{0pt}{3.5ex}  Wavelength (nm) & Mean retardance ($^\circ$) & Mean error ($^\circ$) & RMS error ($^\circ$)  & Std. dev. ($^\circ$)  \\
\hline
\rule[-1ex]{0pt}{3.5ex}  550 & 168.3 & -11.7 & 11.7 & 0.8 \\
\hline
\rule[-1ex]{0pt}{3.5ex}  600 & 178.4 & -1.6 & 1.8 & 0.8 \\
\hline
\rule[-1ex]{0pt}{3.5ex}  650 & 179.1 & -0.9 & 1.1 & 0.6 \\
\hline
\rule[-1ex]{0pt}{3.5ex}  700 & 175.2 & -4.8 & 4.8 & 0.7 \\
\hline
\end{tabular}
\end{center}
\end{table}

\subsection{Known sources of stellar leakage}\label{sec:errors}

From the MMSP measurements, we can estimate the contribution of two sources of leaked starlight in the coronagraph: (1)~fast-axis orientation errors and (2)~retardance errors. To do so, we recalculate the Jones matrices in Eqn.~\ref{eqn:jones1} substituting the actual fast axis angle $\chi^\prime = \chi + \epsilon_\chi$, where $\epsilon_\chi$ is the error in the fast axis angle, and assuming a retardance of $\pi+\epsilon_r$ radians, where $\epsilon_r$ is the retardance error. Thus, the Jones matrix in the linear polarization basis is
\begin{equation}
\mathbf{M}=\mathbf{R}(\chi^\prime)
\left[ \begin{matrix}
   1 & 0  \\
   0 & e^{i(\pi+\epsilon_r)}  \\
\end{matrix} \right]
\mathbf{R}(-\chi^\prime).
\end{equation}
Converting to the circular polarization basis as in Eqn~\ref{eqn:jones1}: 
\begin{equation}
\mathbf{M}_\circlearrowright
=c_V\left[ \begin{matrix}
   0 & e^{i2\chi^\prime}  \\
   e^{-i2\chi^\prime} & 0  \\
\end{matrix} \right] + 
c_L\left[ \begin{matrix}
   1 & 0  \\
   0 & 1  \\
\end{matrix} \right],
\label{eqn:jones2}
\end{equation}
where $c_V$ and $c_L$ are generally functions of position on the mask and wavelength. The fast axis orientation error changes the effective phase shift of the vortex mask to $\exp(\pm i(l\theta + 2\epsilon_\chi))$. The functions $c_V$ and $c_L$ are determined by the retardance error: $c_V=(1+\exp(i\epsilon_r))/2$ and $c_L=(1-\exp(i\epsilon_r))/2$. Other transmission errors, such as dust or debris on the mask, would introduce an additional multiplicative term to $c_V$ and $c_L$ that also potentially depends on position and wavelength. 

The second term in Eqn.~\ref{eqn:jones2}, referred to as the ``leakage term," results in a portion of the transmitted light that doesn't have the vortex phase pattern (i.e. the transmission is the identity matrix for all positions) and whose intensity is $|c_L|^2$ times that of the incident beam, where $|c_L|^2=\sin^2(\epsilon_r/2)$. The fraction of the total power that transfers into the first term, known as the ``vortex term," is $|c_V|^2=\cos^2(\epsilon_r/2)$. However, due to the propagation properties of the vortex term, it is diffracted out of the downstream pupil such that the light is mostly blocked by an aperture stop, known as the Lyot stop (LS). On the other hand, a mean retardance error of $\epsilon_r$~=~1$^\circ$ leads to a leakage term with $|c_L|^2=7.6\times 10^{-5}$. The residual light after the coronagraph in this scenario is an Airy diffraction pattern in the image plane centered on the position of the star. Since the desired dark zone is off-axis with respect to the star, it overlaps with the Airy rings meaning that the raw contrast is $\sim$100$\times$ smaller than the value of $|c_L|^2$. Nonetheless, the retardance errors are too high for our application given the measured retardance errors in Table~\ref{tab:ret_err}.

To cancel the leakage term, we use a polarization filtering method\cite{Serabyn2013} that requires a single circular polarization to be incident on the mask. Since the vortex term transfers to the opposite polarization state and the leakage term remains in the original polarization state, a circular analyzer after the mask reduces the amplitude of the leakage term by its attenuation factor, $\gamma$, which needs to be on the order of $10^2$ (i.e. $10^4$ extinction of the intensity) to ensure the leakage term appears at $<10^{-10}$ contrast.  

In addition to the polarization leakage term and the fast axis orientation error, any amplitude errors due to dust, debris, and other defects in the liquid crystal, adhesives, or on the substrates cause additional stellar leakage. As does the spatial variation of the retardance error. 

To summarize, the important error terms are: 
\begin{enumerate}
    \item Polarization leakage due to global retardance error. Proportional to $|c_L|^2/\gamma^2$. 
    \item Phase pattern error due to errors in the fast axis orientation, $\epsilon_\chi$.
    \item Amplitude errors due to localized defects. Multiplies $c_V$ and $c_L$. 
    \item Amplitude and phase errors due to retardance non-uniformity. Causes $c_V$ and $c_L$ to vary as a function of position. 
\end{enumerate}

Below, we will compare our expected contrast due to the measured mask errors to our testbed data. 

\section{Testbed description} 

\begin{figure}[t]
    \centering
    \includegraphics[width=\linewidth]{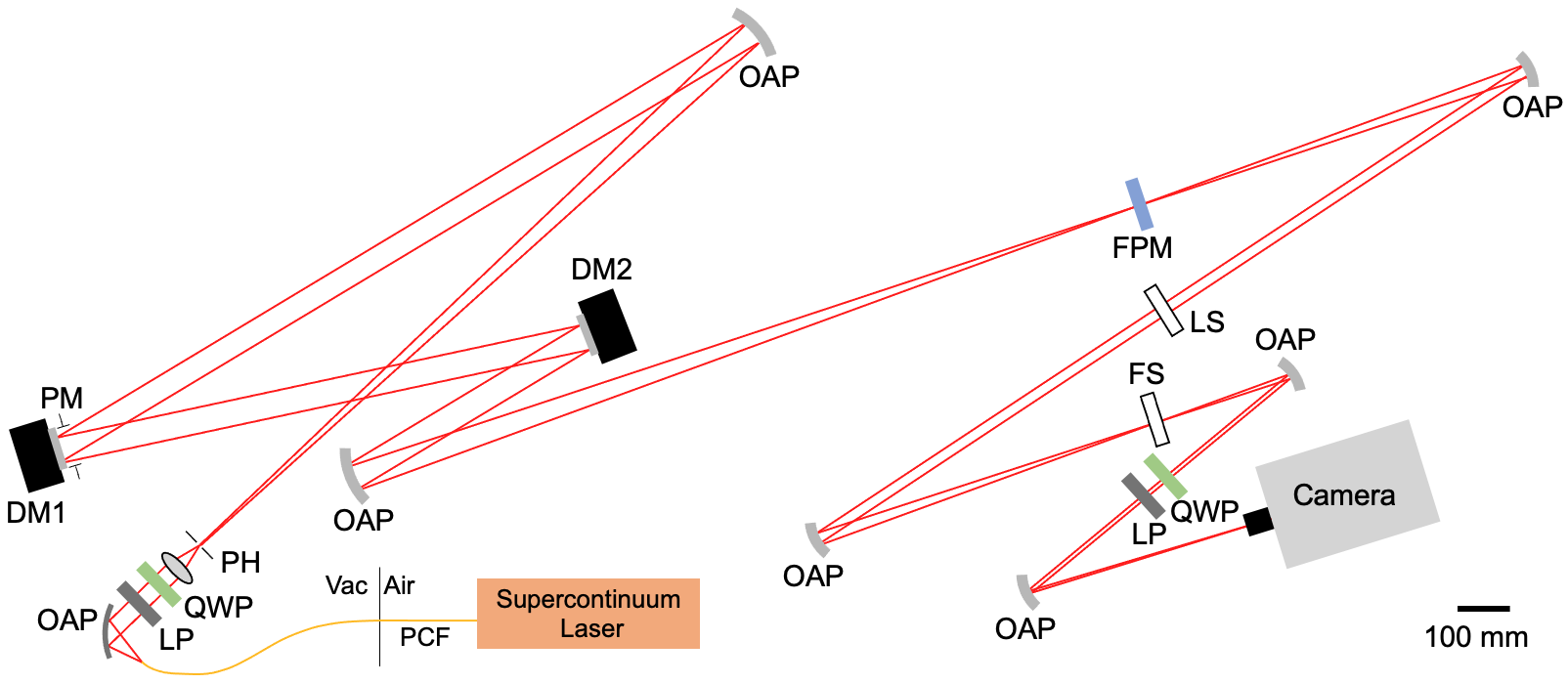}
    \caption{Schematic of the Decadal Survey Testbed (DST). PCF: Photonic Crystal Fiber. Vac: Vacuum. OAP: Off-axis parabolic mirror. LP: Linear Polarizer. QWP: Quarter-wave plate. PH: Pinhole. PM: Pupil mask. DM: Deformable Mirror. FPM: Focal plane mask. LS: Lyot stop. FS: Field stop.}
    \label{fig:dst_layout}
\end{figure}

We tested the vortex mask described above on the Decadal Survey Testbed (DST) in the High Contrast Imaging Testbed (HCIT) facility at NASA's Jet Propulsion Laboratory. The DST is a coronagraph instrument designed to achieve extremely high contrast (on the order of $10^{-10}$)\cite{Patterson2019,Seo2019,Ruane2020_JATIS}. 

Figure~\ref{fig:dst_layout} shows the layout of the DST for this experiment. To simulate the star, a supercontinuum laser source (NKT SuperK Extreme) is coupled to a photonic crystal fiber (PCF) that passes through the air-to-vacuum interface. Since we wish to use polarization filtering, we collimate the light exiting the fiber using an off-axis parabolic mirror (OAP) and create a circular polarization state by passing the beam through a linear polarizer (LP) and quarter wave plate (QWP) with a relative clocking angle of 45$^\circ$. Then the light is focused onto a 4~$\mu$m diameter pinhole (PH) with an achromatic doublet lens to create a pseudo point source representing a star. The light from the PH is collimated by an 1.52~m focal length OAP towards the first deformable mirror (DM1) where a circular pupil mask (PM; 46.6~mm diameter) defines the entrance pupil of the imaging system. The effective resolution is 18~$\mu$m at a wavelength of 550~nm making the pinhole unresolved with a diameter that is 20\% of the angular resolution. A second deformable mirror (DM2) is located 1~m from DM1. DM1 and DM2 have 48$\times$48 actuators based on Lead Magnesium Niobate (PMN) electroceramic arrays with 1~mm pitch manufactured by Northrop Grumman's AOA Xinetics (AOX). After DM2, a second 1.52~m focal length OAP focuses the light onto the focal plane mask (FPM) with a focal ratio of 32.7. The beam transmits through the FPM and a 761~mm focal length OAP creates an image of the pupil with a diameter of 23.3~mm at the position of the Lyot stop (LS). The LS is a circular aperture with a diameter of 18.6~mm (80\% of the geometric pupil diameter). After the beam passes through the Lyot stop, it is focused onto a field stop (FS) in the focal plane that has openings designed to match the dark zone size and shape such that the bright field outside of the intended dark zone is blocked from reaching the camera. After the FS, the light is collimated one last time by an OAP and passes through the circular analyzer (QWP+LP) and is focused by a final OAP onto the camera (Andor Neo sCMOS). 

Several aspects of the DST are designed to maximize wavefront stability. The DST is located in a vacuum chamber in an environmentally-controlled and vibrationally-isolated clean room. The vacuum chamber is at a constant pressure of 0.1~mTorr during operation. The OAPs are bonded into custom made titanium mounts on a carbon fiber table (supplied by CarbonVision GmbH) to minimize thermal expansion. The table is set on passive mechanical isolators (supplied by Minus K Technology) to further reduce vibrations. The DMs have custom-made controllers with high voltage resolution and stability\cite{Bendek2020}. 

Remote motion control is enabled where required. The FPM, LS, and FS are on three-axis stages with micron-level precision. The source apparatus (containing the OAP collimator, LP, QWP, lens, and PH) can be moved in three axes as a unit. The LPs and QWPs are mounted in axial rotation stages and the analyzer can be completely removed from the beam along one dimension as a unit. The camera is on a single-axis stage along the beam direction with $>$200~mm of travel. 

A remote-controlled variable filter (NKT Varia) selects the source wavelengths. The combination of the variable filter and PCF allow wavelength selection from 450~nm to 800~nm. 

There are two insertable lenses: one directly after the pinhole and one just before the camera. Both create an image of the pupil at the plane of the science camera. However, the former is intended to provide a means to evenly illuminate the FPM, which is referred to as the ``diffuser lens." The latter creates a pupil image on the camera for calibration purposes and is therefore called the pupil imaging lens. 

The temperature of the bench is controlled to within $\sim$1~mK at 12 locations, including eight on the table (at 22$^\circ$C), one on each DM (at 21.6$^\circ$C), and one on the camera (with a floating set-point). Away from the control points, the typical peak-to-valley temperature drift on the bench is $\sim$10~mK per 24~hr. The cameras and high voltage DM electronics are also liquid cooled with water to remove excess heat from the testbed. The temperature of the water typically varies significantly more than 10~mK, but these variations do not have a major impact on our performance due low thermal conductivity between the water lines and the bench. 

\section{Procedure} 

\subsection{Testbed alignment and calibration}

Before testing coronagraph masks on DST, we correct the wavefront at the science camera. While the DM voltages required to flatten their surfaces were determined prior to their installation on the testbed, their shapes change over time, especially after venting and pumping down the vacuum chamber. Thus, we first refocus the testbed by moving the camera (and the source position when necessary) along the beam direction with the DMs at their most recent corrected shape. The optimal camera position, determined by the minimum full-width at half-maximum (FWHM) of the point spread function (PSF), can move by 10s of mm between subsequent vacuum cycles. After removing the defocus, we minimize the remaining wavefront error using a phase retrieval method and applying the corrections needed to flatten the wavefront at the science camera. 

\begin{figure}[t]
    \centering
    \includegraphics[width=\linewidth]{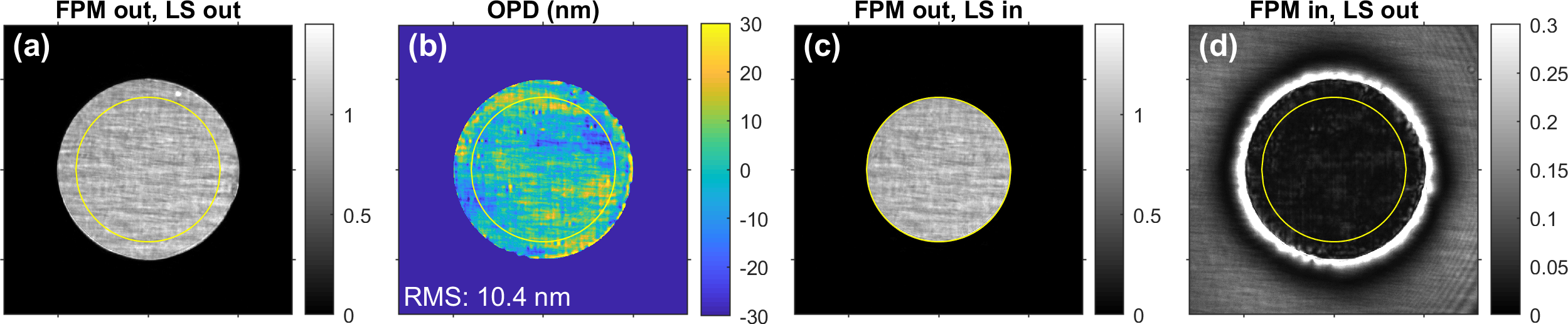}
\caption{Testbed pupil data. \textbf{(a)}~The pupil image with the FPM and LS out of the beam. \textbf{(b)}~Optical path difference (OPD) determined via phase retrieval. The RMS OPD error was 10.4~nm after the wavefront flattening routine. The full pupil diameter is 460~pixels across. \textbf{(c)}~Same as (a), but with the LS in. \textbf{(d)}~Same as (a), but with the FPM in. (a),(c), and (d) are normalized by the median intensity within the beam in (a). The yellow circle shows the relative size of the LS in each case. }
    \label{fig:pupils}
\end{figure}

The wavefront corrections require calibration of the DM. Using phase retrieval measurements, we poke isolated actuators on each DM and determine the DM location, scaling, and rotation with respect to the beam as well as the conversion between voltage and surface motion and the influence function of each actuator\cite{Bendek2020}. Figure~\ref{fig:pupils}a-b show the measured pupil intensity and phase after flattening the wavefront with an OPD residual of 10.4~nm~RMS.  

The Lyot stop is centered on the beam automatically using an algorithm that balances the intensity in the four quadrants of the pupil image using the pupil imaging lens (and with the diffuser lens removed) with the origin defined as the center of the full beam. The resulting aligned image is shown in Fig.~\ref{fig:pupils}c.

The polarization analyzer is aligned automatically by finding the QWP and LP rotation angles that minimizes the intensity on the detector with the FPM totally removed from the beam. We measure the final extinction by taking the ratio between the intensity at the science camera with the beam passing through the FPM in a region away from the phase singularity (effectively passing through a half-wave plate) and the intensity with the FPM totally removed. For the specific optics used in this experiment, the extinction was measured to be 7.5$\times10^{3}$.

The FPM is placed at the beam focus by inserting the diffuser lens just after of the source pinhole, which evenly illuminates the focal plane mask with coherent light. In this configuration, the phase singularity at the center of the FPM appears to create a dark core diffraction pattern if it is not conjugated to the camera. To align the FPM, we scan the position of the FPM along the beam and find the position that minimizes the size of the diffracted core around the phase singularity.  

The FPM is aligned automatically in the dimensions perpendicular to the beam direction by balancing the intensity in the pupil image (Fig.~\ref{fig:pupils}d), as in the case of the Lyot stop alignment, but with the FPM aligned to the beam and comparing the counts in an annulus with inner and outer radii respectively 1.3 and 1.8 times the full pupil radius. 

Once the coronagraph masks are aligned, we remove the lenses and image the off-axis PSF at the desired wavelengths with the FPM offset by 2~mm. We use these PSF images to determine the in-focus beam position in the image and the PSF peak (in units of counts per second) versus wavelength for normalizing the image contrast. 

The final calibration is to determine the plate scale, or the relationship between the spatial dimension at the camera image plane to the angular resolution of the imaging system. To measure this, we apply a sine wave voltage pattern to DM1 and determine the location of the corresponding PSF copies created in the image plane. Based on the previously determined number of actuators across the beam, we infer the number of pixels per $\lambda_0/D$.

\subsection{Focal plane wavefront sensing and control}

\begin{figure}[t]
    \centering
    \includegraphics[width=0.9\linewidth]{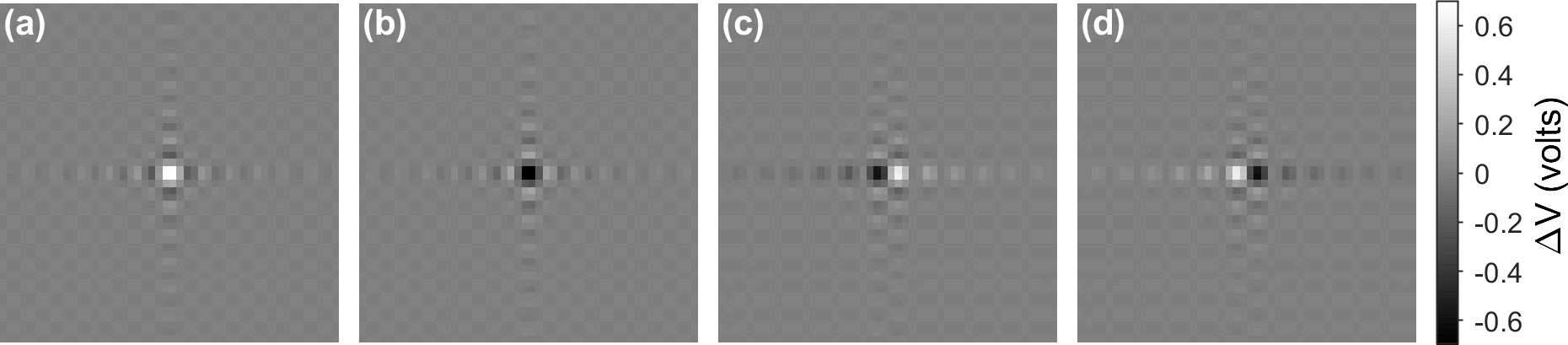}
    \caption{Voltage changes (``probes") applied to DM1 to estimate the electric field in the correction region. (a),(b) and (c),(d) are pairs with the opposite parity. }
    \label{fig:probeVoltages}
\end{figure}

After the initial calibrations and wavefront correction, we choose the region of the image and wavelength range over which we wish to create the dark zone. Then we run an iterative process that estimates the E-field in the image plane and determines the DM voltages to apply in order to cancel the E-field. Our testbed implementation is available as part of the open-source Fast Linearized Coronagraph Optimizer (FALCO) software package\cite{Riggs2018}. 

\subsubsection{E-field sensing via pair-wise probing}

The focal plane E-field estimation method is known as pair-wise probing (PWP)\cite{Giveon2011,Groff2015}, which applies sequential DM surface changes that are intended to evenly illuminate the dark zone and make a phase-shifting interferometric measurement at each pixel in the dark zone simultaneously. We use two pairs of DM surface probes of the form:
\begin{equation}
\begin{split}
    \Delta h_{1,\pm} &= \pm A\;\text{sinc}(x)\;\text{sinc}(y)\;\cos(y),\\
    \Delta h_{2,\pm} &= \pm A\;\text{sinc}(x)\;\text{sinc}(y)\;\sin(y),
\end{split}
\end{equation}
where $A$ is the probe amplitude and the origin assumed to be the center of the DM (see Fig.~\ref{fig:probeVoltages}). These probe shapes modulate a rectangular region of the image plane. At each iteration and control passband, we take four images with each of the DM probes applied as well as the un-probed image. The E-field estimate is computed anew at each iteration as a batch process using the linearized least-squares approach described by Give'on et al.\cite{Giveon2011}.  

\begin{figure}[t]
    \centering
    \includegraphics[width=0.8\linewidth]{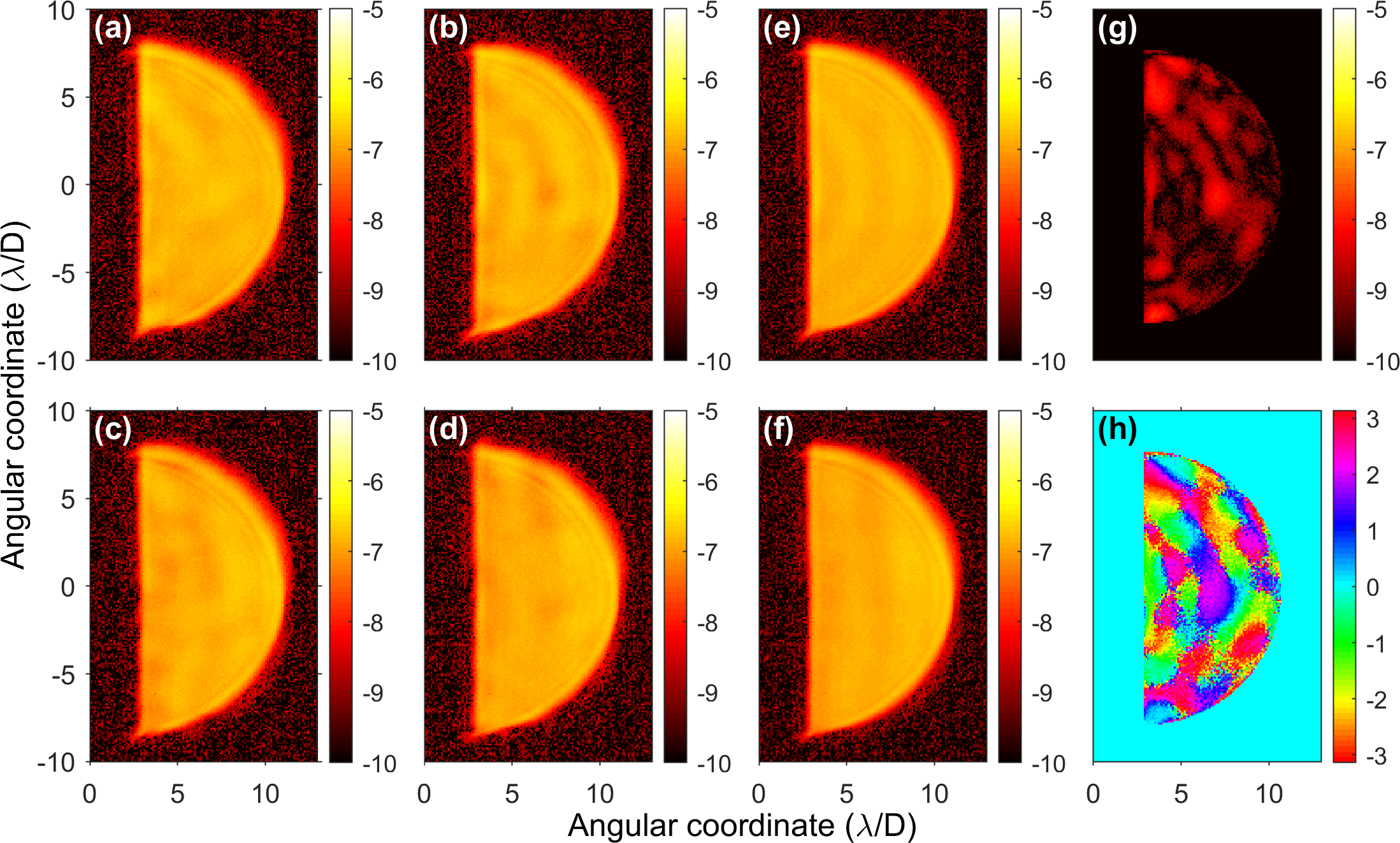}
    \caption{E-field estimation example using pair-wise probing. \textbf{(a)-(d)}~Raw normalized intensity images with probe with the corresponding probe patterns in Fig.~\ref{fig:probeVoltages} applied to the DM. \textbf{(e)}~Computed probe intensity, $|\Delta P|^2$, for the (a)-(b) probe pair. \textbf{(f)}~Same as (e), but for the (c)-(d) probe pair. \textbf{(g)-(h)}~Estimated E-field in the correction region. (g) is the modulated component of the normalized intensity and (h) is the phase. While the probe patterns address a rectangular region of the image, the `D' shape boundary is defined by the FS opening.}
    \label{fig:estimationExample}
\end{figure}

Figure~\ref{fig:estimationExample} shows an example estimation in the dark zone. The images for each probe pair, $I_{1,\pm}$ (Fig.~\ref{fig:estimationExample}ab) and $I_{2,\pm}$ (Fig.~\ref{fig:estimationExample}cd), are used to compute the probe intensity in each case (Fig.~\ref{fig:estimationExample}ef): 
\begin{equation} 
    \begin{split}
        |\Delta P_1|^2 &= \frac{(I_{1,+})+(I_{1,-})}{2} - I_0,\\
        |\Delta P_2|^2 &= \frac{(I_{2,+})+(I_{2,-})}{2} - I_0,
    \end{split}
\end{equation}
where $I_0$ is the unprobed image. The phase of the probes is approximated using a simplified numerical propagation model of the coronagraph, known as the ``compact" model.

The compact model contains information representing the optical system. The model treats the aberrations measured via phase retrieval as occurring in a single input pupil plane and computes the propagations between subsequent pupil and focal planes using Fourier transforms and uses the Fresnel approximation to propagate between the DMs. The model also includes mask designs and calibration data such as the DM registration with respect to the beam, the conversion between the DM voltage and surface displacement, the basic FPM and LS design, the chosen dark zone, and the control wavelengths. With this information, we can predict the change in E-field for a given voltage change at the DM, which is what we need to determine the focal plane E-field as well as to carry out the control step described below. 

\subsubsection{Electric field conjugation}

After sensing the E-field in the dark zone, the broadband electric field conjugation (EFC) algorithm\cite{Giveon2009,Groff2015} is used to determine the DM voltages needed to conjugate the measured E-field. Using the compact model, we determine the linearized interaction between the complex E-field and each DM actuator and build a matrix with the vectorized dark zone pixels as the columns. This interaction matrix is also known as the Jacobian, and is the derivative of the E-field with respect to DM voltages. Our implementation ignores actuators that have a weak influence in the dark zone, which typically corresponds to those outside of the beam footprint on the DM. In each control step, the DM commands are computed as the Tikhonov-regularized left pseudoinverse of the interaction matrix times the most recently estimated E-field.  

\begin{figure}[t]
    \centering
    \includegraphics[width=0.75\linewidth]{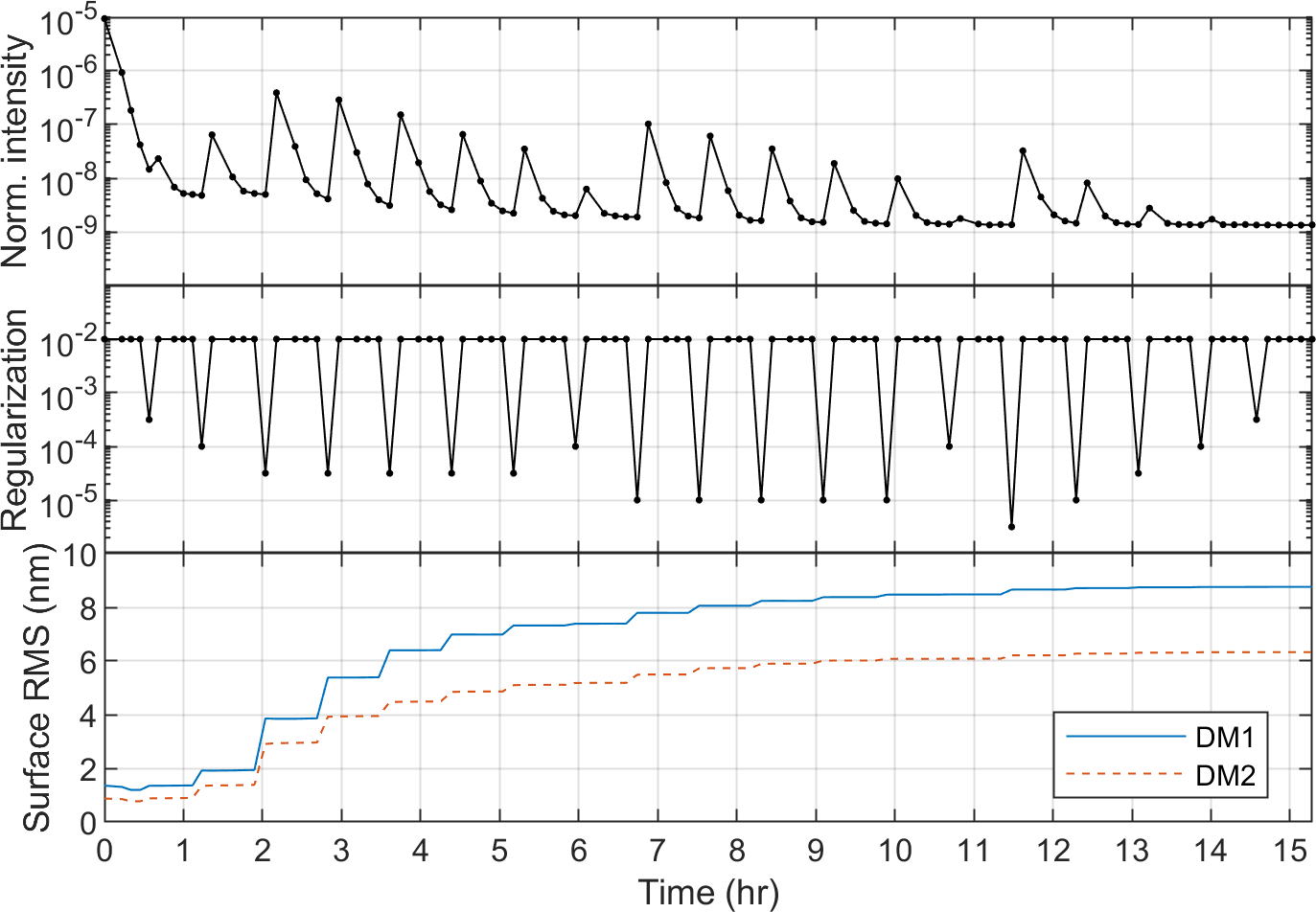}
    \caption{An example EFC convergence plot showing the \textbf{(top)} spatially- and spectrally-averaged raw normalized intensity, \textbf{(middle)} regularization parameter, and \textbf{(bottom)} RMS surface applied to each DM. }
    \label{fig:EFCconvergence}
\end{figure}

Finding the best choice for the regularization parameter requires trial-and-error for a given coronagraph system. We find that the ``beta-bumping" technique of Seo et al.\cite{Seo2017} is the most effective approach where a weak regularization is used for most iterations, but a periodic aggressive regularization step is scheduled in advance. Figure~\ref{fig:EFCconvergence} shows an example of the sensing and control process. In our case, the aggressive regularization is used every five iterations and gets progressively stronger. After an aggressive regularization step, the intensity in the dark hole jumps to a worse value and the DM moves more significantly than at other iterations. Thus, we also re-linearize (i.e.~re-compute the Jacobian matrix) only after the aggressive steps to save computational time. 

Figure~\ref{fig:EFCconvergence} also highlights two common regimes we encounter when working at high contrast. The first $\sim$10 iterations are correcting so-called ``easy" dark zone modes, which leads to quick convergence (e.g. we reach $\sim5\times10^{-9}$ in $\sim$1~hr of testbed time), but the contrast generally reaches a floor. The second regime is where the beta-bumping allows us to break through this initial floor. This occurs for several possible reasons including that the control is addressing higher order modes in the dark zone or that the algorithm is otherwise stuck due to model mismatch or in a local minimum. The beta-bumping scheme can help in any of these scenarios. However, improving upon on the initial dark zone using beta-bumping can be much more time consuming; our example in Fig.~\ref{fig:EFCconvergence} shows improvement for several more hours after the initial floor.  

\section{Results} 

\begin{figure}[t]
    \centering
    \includegraphics[width=0.75\linewidth]{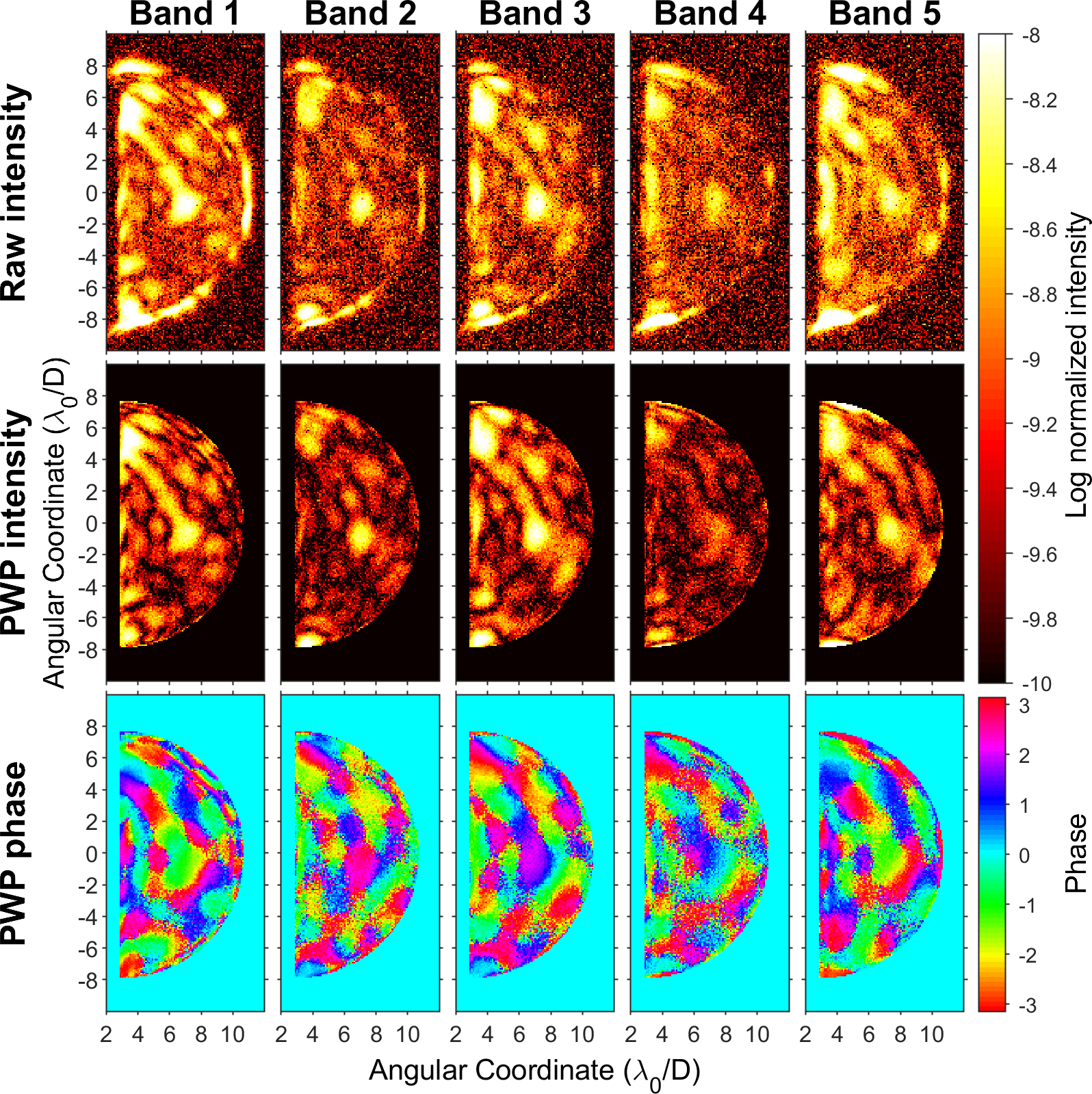} 
    \caption{Images after focal plane wavefront sensing and control. \textbf{(top row)}~The raw normalized intensity in each sub-band in Table~\ref{tab:normIbyBand}, also referred to as the un-probed images. \textbf{(middle row)}~The intensity as determined from the PWP estimation process. \textbf{(bottom row)}~The phase of the estimated E-field via PWP.}
    \label{fig:DHgrid}
\end{figure}

\begin{figure}[t]
    \centering
    \includegraphics[height=0.4\linewidth]{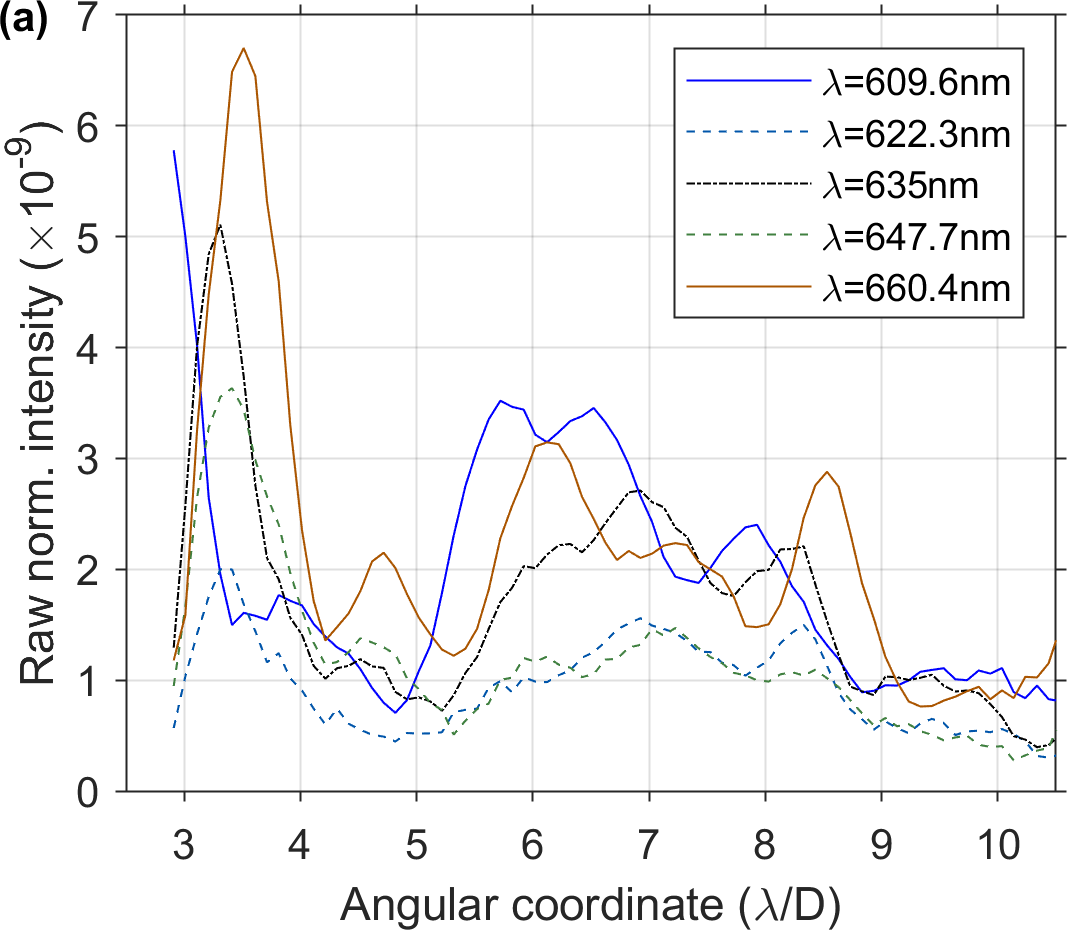}
    \includegraphics[height=0.4\linewidth]{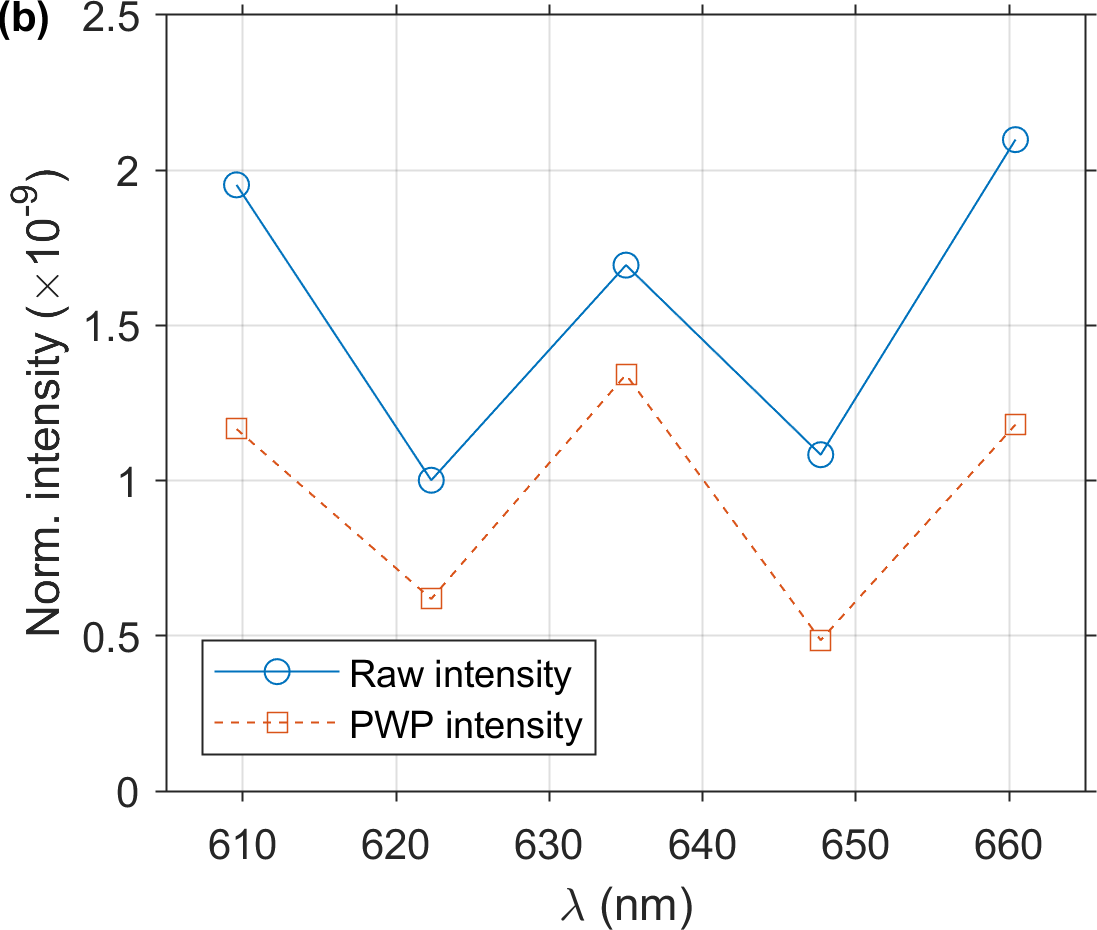}
    \caption{Normalized intensity in five sub-bandpasses for a dark zone with a total fractional bandwidth of $\Delta\lambda_0/\lambda_0$~=~0.1. \textbf{(a)}~Azimuthal average of the raw normalized intensity. \textbf{(b)}~Spatial average of raw and PWP intensity in each sub-bandpass.  } 
    \label{fig:10pctresult}
\end{figure}

Our best broadband contrast to date with a vector vortex coronagraph was achieved during a testbed run carried out on December 18-22, 2021. After the full wavefront sensing and control procedure shown in Fig.~\ref{fig:EFCconvergence}, we achieved mean normalized intensity of 1.6$\times10^{-9}$ in a dark zone defined over 3-10 $\lambda_0/D$ and five wavelength bands that make up 603.3-666.8~nm (a 10\% bandwidth centered at $\lambda_0$~=~635~nm). The results are shown in Figs.~\ref{fig:DHgrid} and \ref{fig:10pctresult} and the numerical results are detailed in Table~\ref{tab:normIbyBand}. 

\begin{table}[t] 
\caption{Mean normalized intensity in dark zone (3-10 $\lambda_0/D$) at each sub-band over a 10\% bandwidth after convergence.} 
\label{tab:normIbyBand}
\begin{center}       
\begin{tabular}{|c|c|c|c|} 
\hline
\rule[-1ex]{0pt}{3.5ex}  Band & Wavelength range (nm) & Raw intensity & PWP intensity  \\
\hline
\rule[-1ex]{0pt}{3.5ex}  1 & 603.3 - 616.0 & 1.95$\times10^{-9}$ & 1.17$\times10^{-9}$ \\
\hline
\rule[-1ex]{0pt}{3.5ex}  2 & 616.0 - 628.7 & 1.00$\times10^{-9}$ & 6.19$\times10^{-10}$ \\
\hline
\rule[-1ex]{0pt}{3.5ex}  3 & 628.7 - 641.4 & 1.69$\times10^{-9}$ & 1.34$\times10^{-9}$ \\
\hline
\rule[-1ex]{0pt}{3.5ex}  4 & 641.4 - 654.1 & 1.08$\times10^{-9}$ & 4.86$\times10^{-10}$ \\
\hline
\rule[-1ex]{0pt}{3.5ex}  5 & 654.1 - 666.8 & 2.10$\times10^{-9}$ & 1.18$\times10^{-9}$ \\
\hline
\rule[-1ex]{0pt}{3.5ex}  All & 603.3 - 666.8 & 1.57$\times10^{-9}$ & 9.59$\times10^{-10}$ \\
\hline
\end{tabular}
\end{center}
\end{table} 

\begin{figure}[t]
    \centering
    \includegraphics[height=0.4\linewidth]{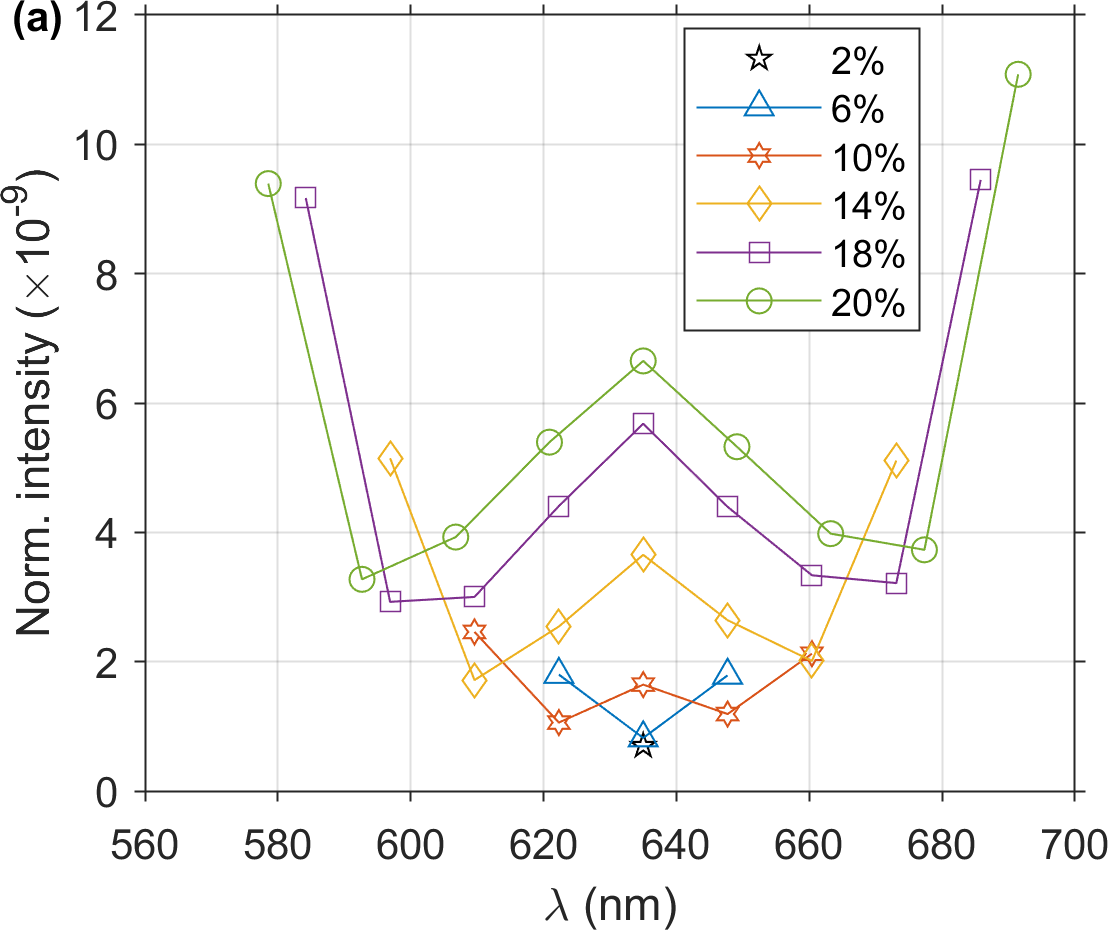}
    \includegraphics[height=0.4\linewidth]{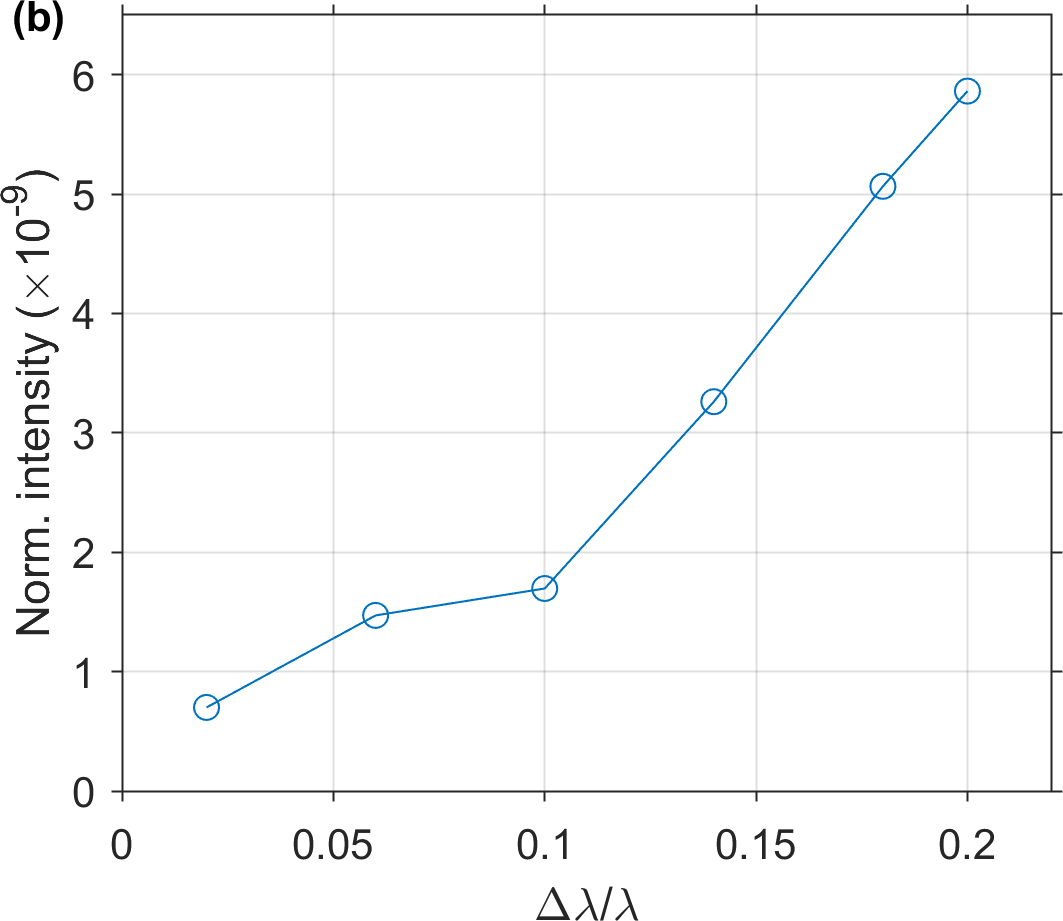}
    \caption{\textbf{(a)}~Mean raw normalized intensity in 3-10 $\lambda/D$ dark zone versus percent bandwidth at the control wavelengths. \textbf{(b)}~The spectral mean of (a) versus $\Delta\lambda_0/\lambda_0$. In each case, the focal plane wavefront sensing and control process was repeated to find the best DM settings for minimized the normalized intensity in each bandwidth.} 
    \label{fig:normIvsBW}
\end{figure}

We also investigated the degradation as a function of spectral bandwidth. Repeating the focal plane wavefront sensing and control procedure above, we started with a single 2\% band, then increased the bandwidth by adding two 2\% bands on each end of the full spectral range up until 18\% (using nine sub-bands). Finally, we added a 20\% case with nine sub-bands since future mission concepts have baselined 20\% bandwidths\cite{HabEx_finalReport}. Figure~\ref{fig:normIvsBW} shows the resulting intensity versus wavelength for each case as well as the dependence of the mean intensity on the total bandwidth. For the 20\% bandwidth case, the mean normalized intensity was 5.9$\times10^{-9}$. 

\section{Discussion} 

From the 10\% bandwidth case, we can determine the current contrast limitations for vector vortex coronagraphs. The most important outcome is that the raw intensity in Fig.~\ref{fig:DHgrid} is similar in magnitude and morphology to the intensity determined via PWP. This means that the dominant intensity source in the dark hole is coherent. However, the intensity of the coherent light is changing in an unusually chromatic way. We attribute this to features on the focal plane mask that create bright speckles in the image. While the DMs can create Fourier modes that place bright speckles of arbitrary amplitude and phase in the dark zone, the bright speckle must move outward with wavelength due to diffraction. On the other hand, a bright speckle due to a FPM feature is fixed in position in the image plane and therefore can only be perfectly corrected at one or two wavelengths depending on the total bandwidth and number of DMs used. As described in Section~\ref{sec:errors}, such FPM features include spatially-varying transmission errors, retardance errors, fast-axis orientation errors, as well as other phase errors. 

The polarization leakage creates an incoherent Airy pattern in the image plane that can be measured directly from the intensity in the dark zone using monochromatic laser light. In doing so, we determined that the polarization leakage has a mean intensity of $4\times10^{-10}$. The only remaining term in the error budget that leads to coherent residuals is the quantization error in the DM electronics\cite{Ruane2020_LSB}, but we estimate this to be on the level of $\sim1\times10^{-10}$ for our DMs and controller electronics (and to be improved using a next generation of DM electronics; see Bendek et al., these proceedings). The majority of the remaining error budget (accounting for $\ge 6\times10^{-10}$) is associated with FPM errors. 

To add confidence to this conclusion, we performed an additional experiment where we removed the vortex FPM and circular analyzer and introduced a simple Lyot coronagraph made up of a circular Ni occultor on a glass substrate (as used for DST commissioning\cite{Seo2019}). With all else the same, we were able to achieve $4\times10^{-10}$ total contrast in the same dark zone. This provides further evidence that the dominant source of coherent light in the dark zone in the case of the vortex FPM is the spatially-varying imperfections in the mask itself.  

When dominated by these coherent speckle effects, we find that the mean normalized intensity scales roughly with the square of the bandwidth (Fig.~\ref{fig:normIvsBW}). On the contrary, when limited by polarization leakage, the intensity is less dependent on bandwidth assuming the wavelengths of interest are within the regime where the HWP that makes up the vortex FPM and the QWPs are achromatic. 

\section{Conclusion} 

We have presented new laboratory results for the liquid crystal polymer based vector vortex coronagraph technology. The mean normalized intensity averaged over 3-10~$\lambda_0/D$ separations on one side of the pseudo-star is 1.6$\times10^{-9}$ in a 10\% bandwidth and 5.9$\times10^{-9}$ in 20\% bandwidth. We showed that the residual stellar intensity is coherent and likely dominated by spatially-varying imperfections in the vortex focal plane mask. As a result, the normalized intensity scales roughly with the square of the bandwidth for bandwidths $>$10\%. Our current project aims to achieve 5$\times10^{-10}$ in a 10\% bandwidth using a new generation of vector vortex masks with tighter tolerances on their imperfections and defects. We also plan to improve the suppression of the polarization leakage and switch to a new set of higher-resolution DM controller electronics to reduce quantization errors.

\acknowledgments 
 
The authors thank David Roberts (Beam Co.) for useful comments. The research was carried out at the Jet Propulsion Laboratory, California Institute of Technology, under a contract with the National Aeronautics and Space Administration (80NM0018D0004).

\bibliography{report} 
\bibliographystyle{spiebib} 

\end{document}